\documentclass[twocolumn,showpacs,amsmath,amssymb,floatfix,prl]{revtex4}
\usepackage{graphicx}
\usepackage{dcolumn}
\usepackage{bm}

\begin{document}

\title{
Many-particle localization by constructed disorder: enabling quantum
computing with perpetually coupled qubits}

\author{M.I.~Dykman$^{1}$, F.M.~Izrailev$^{3}$, L.F. Santos$^{1}$, and
M.~Shapiro$^{2}$}
\affiliation{ $^{1}$Department of Physics and
Astronomy, and $^{2}$Department of Mathematics, Michigan State
University, East Lansing, MI 48824\\
$^{3}$ Instituto de F\'isica,
Universidad Aut\'onoma de Puebla, Puebla 72570, M\'exico }

\date{\today}

\begin{abstract}
We demonstrate that, in a many-particle system, particles can be
strongly confined to their sites. The localization is obtained by
constructing a sequence of on-site energies that efficiently
suppresses resonant hopping. The time during which a many-particle
state remains strongly localized in an infinite chain can exceed the
reciprocal hopping frequency by $\agt 10^5$ already for a moderate
energy bandwidth. The results show viability of quantum computers with
time-independent qubit coupling.
\end{abstract}


\maketitle

Disorder-induced localization has been one of the central problems of
condensed-matter physics \cite{Anderson}. The problem is particularly
challenging for many-body systems, where only a limited number of
results has been obtained \cite{Altshuler03}.  Recently it has also
attracted much interest in the context of quantum computing. In many
proposed physical implementations of a quantum computer (QC) the
qubit-qubit interaction is not turned off
\cite{liquid_NMR,Makhlin01,mark,mooij99,Yamamoto02,van_der_Wiel03}.
Generally, the interaction leads to hopping of excitations between the
qubits. Preventing hopping is a prerequisite for quantum
computation. To control a QC and measure its state excitations must
remain localized between operations. Several approaches to quantum
computation with interacting qubits were proposed recently
\cite{Zhou02,Benjamin03,Berman-01}.

In this paper we study strong on-site many-particle localization. It
implies that each particle (or excitation) is nearly completely
confined to one site. This is a stronger condition than just
exponential decay of the wave function, and it is this condition that
must be met in a QC.

Strong on-site localization does not arise in a disordered
many-particle system with bounded random on-site energies
\cite{Shepelyanskiy}. Indeed, consider a state where particles occupy
$N$ sites. For short-range hopping it is directly coupled to $\sim N$
other $N$-particle states. With probability $\propto N$ one of them
will be in resonance with the initial state, provided the on-site
energies are uniformly distributed over a finite-width band. For large
$N$ this leads to state hybridization over time $\sim
J^{-1}$, where $J$ is the intersite hopping integral (we set $\hbar =
1$).

Here we show that many-particle localization can be obtained in an
{\em infinite} chain by {\em constructing} a sequence of on-site
energies. For the proposed narrow-band sequence, all many-particle
states remain confined for a time that largely exceeds $J^{-1}$. We
find that stationary many-particle states in moderately long chains
are also strongly localized.

A QC allows both studying and using strong localization. Here, on-site
excitation energies are interlevel distances of the qubits. They can
often be individually controlled, which makes it possible to construct
an arbitrary energy sequence. However, since the qubit tuning range is
limited, so should be the energy bandwidth. A smaller bandwidth is
also desirable for a higher speed of quantum gate operations.

To localize one particle, the difference between excitation energies
on neighboring sites should be much larger than $J$.  In addition,
even for nearest neighbor coupling, the energies of remote sites
should also differ to prevent virtual transitions via intermediate
sites. However, the further away the sites are, the smaller their energy
difference can be. We use this to obtain strong on-site
single-particle confinement for a bounded energy bandwidth.

For many-particle localization one has to suppress not only
single-particle, but also combined resonances, where several
interacting excitations make a transition simultaneously. There is no
known way to eliminate all such resonances. However, the problem can
be approached from a different point of view by looking at a lifetime
of a localized state. The effective many-particle hopping integral
quickly falls off with the increasing number of involved excitations
and intermediate nonresonant sites, which gives the effective
``order'' of a transition. To obtain a desired lifetime it is
sufficient to eliminate resonances up to a certain order. As we show,
this can be done for an infinite system.

We will consider a one-dimensional chain of $S=1/2$ spins in a
magnetic field. This model also describes an array of qubits. The
excitation energy of a qubit is the Zeeman energy of a spin. The
qubit-qubit interaction is the exchange spin coupling. For many
proposed realizations of QC's
\cite{liquid_NMR,Makhlin01,mark,mooij99,Yamamoto02,van_der_Wiel03} it
has a form ${1\over 2}\sum^{\prime}
J_{nm}^{\mu\mu}S_n^{\mu}S_m^{\mu}$, where $n,m$ are spin sites,
$\mu=x,y,z$ are spin projections, and $J_{nm}^{xx}=J_{nm}^{yy}$ for
the effective magnetic field in the $z$-direction.  The 1D spin system
can be mapped, via Jordan-Wigner transformation, onto a system of
fermions. For nearest neighbor coupling, the fermion Hamiltonian is
\begin{eqnarray}
\label{hamiltonian_fermions}
H=&&\sum\nolimits_n\varepsilon_na_n^{\dagger}a_n+ {1\over
2}J\sum\nolimits_n\bigl( a_n^{\dagger}a_{n+1}+a_{n+1}^{\dagger}a_n\bigr)
\nonumber\\ &&+ J\Delta\sum\nolimits_n
a_n^{\dagger}a_{n+1}^{\dagger}a_{n+1}a_n.
\end{eqnarray}
Here, $a^{\dagger}_n, a_n$ are the fermion creation and annihilation
operators. Presence of a fermion on site $n$ corresponds to the $n$th
spin (qubit) being in the excited state. The on-site energies
$\varepsilon_n$ in Eq.~(\ref{hamiltonian_fermions}) are the Zeeman
energies counted off from the characteristic central energy, $J\equiv
J_{n\,n+1}^{xx}$ is the hopping integral, and $J\Delta \equiv
J_{n\,n+1}^{zz}$ is the fermion interaction energy; we set $J,\Delta>
0$.

Localization of stationary states can be conveniently characterized by
the inverse participation ratio (IPR), which shows over how many sites
the wave function spreads. For an $N$-particle eigenstate
$|\psi_{N\lambda}\rangle$ ($\lambda$ enumerates the eigenstates) it is given by
\begin{equation}
\label{IPR}
I_{N\lambda}=\left(\sum\nolimits_{n_1<\ldots<n_N}\bigl\vert\langle
\Phi_{n_1\ldots n_N}|\psi_{N\lambda}\rangle\bigr\vert^4\right)^{-1},
\end{equation}
where $|\Phi_{n_1\ldots n_N}\rangle = a^{\dagger}_{n_1}\ldots
a^{\dagger}_{n_N}|0\rangle$ is an on-site $N$-particle wave function
(quantum register).

For fully localized stationary states $I_{N\lambda}=1$. For
delocalized states $I_{N\lambda}\gg 1$ [for an $L$-site chain
$I_{N\lambda}\lesssim L!/N!(L-N)!$]. Strong localization corresponds
to $I_{N\lambda}$ being close to $1$ for all states.

Localization requires that the on-site energies $\varepsilon_n$ be
tuned away from each other.  For nearest
neighbor coupling a natural first step is to separate
$\varepsilon_n$'s into two subbands, for even and odd $n$, with the
inter-subband distance $h$ that significantly exceeds $J$.  Then we
further split each subband into two subbands to detune next nearest
neighbors. Here the splitting can be smaller, because
next-nearest-neighbor hopping occurs via a nonresonant site, and the
effective hopping integral is $\sim J^2/h$. The procedure of band
splitting is continued, with higher-order splitting being smaller and
smaller.

\begin{figure}[h]
\includegraphics[width=2.8in]{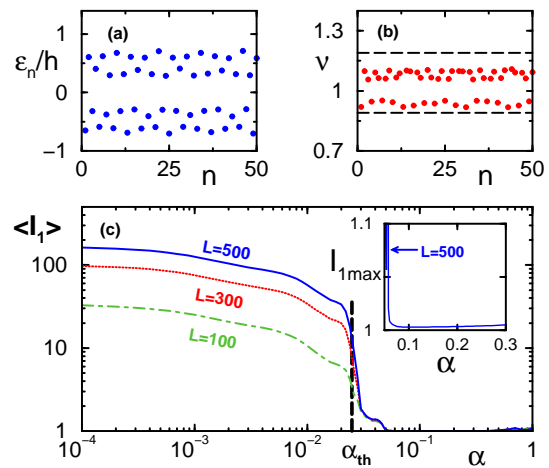}
\caption{(color online). Single-particle localization for the on-site
energy sequence (\protect\ref{sequence}). (a) The energies
$\varepsilon_n/h$ for $\alpha=0.3$. (b) The decrement $\nu$ of the
$\alpha$-dependence of the transition amplitude $K_n(m)$
(\ref{result}) for $m=1000$ as function of $n$. The dashed lines show
the analytical limits on $\nu$.  (c) The mean single-particle inverse
participation ratio $\langle I_1\rangle$ vs. $\alpha$ for $h/J=20$ and
for different chain lengths $L$. The vertical dashed line shows the
analytical estimate for the threshold of strong localization. The
inset shows the maximal IPR, $I_{1\max}=\max_{\lambda} I_{1\lambda}$,
demonstrating strong localization.}
\label{fig:one_excitation}
\end{figure}

A simple sequence of $\varepsilon_n$ that implements the above idea
has the form
\begin{equation}
\label{sequence}
\varepsilon_n={1\over 2}h\left[(-1)^n -\sum\nolimits_{k=2}^{n+1}(-1)^{\lfloor
n/k\rfloor}\alpha^{k-1}\right], \quad n\geq 1
\end{equation}
($\lfloor \cdot\rfloor$ is the integer part). The energies
(\ref{sequence}) are illustrated in Fig.~\ref{fig:one_excitation}(a).
Besides the scaling factor $h$, they are characterized by one
dimensionless parameter $\alpha <1$.  For small $\alpha$, the two
major subbands have width $\approx \alpha h$ and are separated by a
gap of width $\approx h$. The splitting of higher-order subbands are
proportional to higher powers of $\alpha$. For $\alpha \agt 0.4$ all
subbands overlap and the subband structure disappears.

One can see from Eq.~(\ref{sequence}) and
Fig.~\ref{fig:one_excitation}(a) that sites with close energies are
indeed spatially separated. Analytical estimates of the energy
difference can be obtained for small $\alpha$. We have
$|\varepsilon_{n+m}-\varepsilon_n| \sim h$
for odd $m$ and $\sim \alpha h$ for odd $m/2$. In general, the larger
is $m$ the higher may be the order in $\alpha$ of the leading term in
$|\varepsilon_{n+m}-\varepsilon_n|$.

It is important for localization that the sequence (\ref{sequence})
has no simple symmetry. It is neither self-similar nor quasi-periodic
(which is another example of ``constructed'' disorder
\cite{Sokoloff85}). For analytical estimates it is essential that the
coefficients at any given power $\alpha^q$ are repeated with period
$2(q+1)$ \cite{elsewhere}.

A convenient characteristic of the on-site energy sequence is the
amplitude of a particle transition from site $n$ to site
$n+m$. To the lowest order in $J$ it has the form
\begin{equation}
\label{amplitude}
K_n(m)= \prod\nolimits_{k=1}^{m}
J/\left[2(\varepsilon_n - \varepsilon_{n+k})\right].
\end{equation}
It can be shown using some results from number theory that $K_n(m)$
decays with $m$ nearly exponentially \cite{elsewhere}. For small
$\alpha$ and large $|m|$ we have
\begin{equation}
\label{result}
K_n(m)= \alpha^{-\nu|m|}\,(J/2h)^{|m|}.
\end{equation}
The decrement $\nu$ depends on $n,m$. However, it is limited to a
narrow region around $\nu= 1$ with $0.89 < \nu < 1.19$, cf.
Fig.~\ref{fig:one_excitation}(b). For estimates one can use $\nu=1$,
i.e., set $K_n(m)=K^m, K=J/2\alpha h$.

Equation (\ref{result}) describes the tail of the transition amplitude
for $J/2h\alpha\ll 1$. Strong single-particle localization occurs for
$\alpha \gg \alpha_{\rm th}$, where the threshold value of $\alpha$ is
$\alpha_{\rm th} \approx J/2h$. The condition $\alpha_{\rm th}\ll
\alpha < 0.4$ can be satisfied for a moderately large ratio
of the energy bandwidth $h$ to the hopping integral $J$.

Strong single-particle localization for $h/J=20$, as evidenced by
$I_{1\lambda}$ being very close to 1, is seen from
Fig.~\ref{fig:one_excitation}(c). The data are obtained by
diagonalizing the Hamiltonian (\ref{hamiltonian_fermions}) for open
chains with different numbers of sites $L$.

In the limit $\alpha\to 0$ the stationary single-particle states are
sinusoidal, which gives $\langle I_1\rangle \approx L/3$,
cf. Fig.~\ref{fig:one_excitation}(c) ($\langle\cdot\rangle$ means
averaging over the eigenstates). As $\alpha$ increases, the
bands are split into more and more subbands, and $\langle I_1\rangle$
decreases. It sharply drops to $\approx 1$ in a narrow region, which
can be conditionally associated with a smeared transition to strong
localization. The center of the transition region gives $\alpha_{\rm
th}$. It appears to be independent of the chain length $L$. The
estimate $\alpha_{\rm th}=J/2h$ is in good agreement with the
numerical data for different $h/J$.

When $\alpha \gg \alpha_{\rm th}$, all states are localized. The wave
function tails are small and limited mostly to nearest neighbors. At
its minimum over $\alpha$ for given $h/J$, for all states
$I_{1\lambda} - 1 \approx J^2/h^2$, see inset in
Fig.~\ref{fig:one_excitation}(c). The agreement with the above
estimate becomes better with increasing $h/J$. For $\alpha \agt 0.4$,
when the bands of $\varepsilon_n$ start overlapping, the IPR increases
with $\alpha$.

The difference in the localization problems for many-particle and
single-particle systems stems from the interaction term $\propto
J\Delta$ in the Hamiltonian (\ref{hamiltonian_fermions}). If $\Delta =
0$ (the interaction between the underlying spins or qubits is of the
$XY$-type), the above single-particle results apply to the
many-particle system. For nonzero $\Delta$, on the other hand, (i) the
energy levels are shifted depending on the occupation of neighboring
sites, potentially leading to many-particle resonances, and (ii) there
occur interaction-induced many-particle transitions.

To analyze many-particle effects, it is convenient to
change from $a_n^{\dagger},a_n$ to new creation and annihilation
operators $b_n^{\dagger},b_n$ that diagonalize the single-particle
part of the Hamiltonian
(\ref{hamiltonian_fermions}), $a_n=\sum\nolimits_kU_{nk}b_k$. The
interaction part of the Hamiltonian becomes
\begin{equation}
\label{H_int}
H_i=J\Delta\sum
V_{k_1k_2k_3k_4}b_{k_1}^{\dagger}b_{k_2}^{\dagger}b_{k_3}b_{k_4},
\end{equation}
where the sum runs over $k_{1,2,3,4}$, and $V_{k_1k_2k_3k_4} =
\sum\nolimits_pU^*_{pk_1}U^*_{p+1\,k_2}U_{p+1\,k_3}U_{pk_4}$.

If all single-particle stationary states are strongly localized, the
off-diagonal matrix elements $U_{nk}$ are small. They are determined
by the decay of the wave functions and fall off exponentially,
$U_{nk}\sim K^{|k-n|}$ for $|k-n| \gg 1$. At the same time, the
diagonal matrix element is $U_{nn} \approx 1$. Therefore the major
terms in the matrix $V_{k_1k_2k_3k_4}$ are those with $\varkappa=0$,
where
\[\varkappa=\min_p(|k_1-p|+|k_2-p-1|+|k_3-p-1|+|k_4-p|).\]
They lead to an energy shift $\propto J\Delta$ for the many-particle
states with occupied neighboring sites.

The terms $V_{k_1k_2k_3k_4}$ with $\varkappa> 0$ lead to two-particle
intersite transitions $(k_3,k_4)\to (k_1,k_2)$, and $VJ\Delta $ plays
the role of a two-particle hopping integral. The parameter $\varkappa$
gives the number of intermediate steps involved in a
transition. The steps are counted off from the configuration where the
particles occupy neighboring sites \cite{elsewhere}. For large $\varkappa$ and
$\alpha\gg \alpha_{\rm th}$ we have $V\propto
(J/2h\alpha^{\nu})^{\varkappa}\ll 1$, i.e., many-particle hopping
integrals are small and rapidly decrease with the number of involved
virtual steps.

\begin{figure}[h]
\includegraphics[width=2.8in]{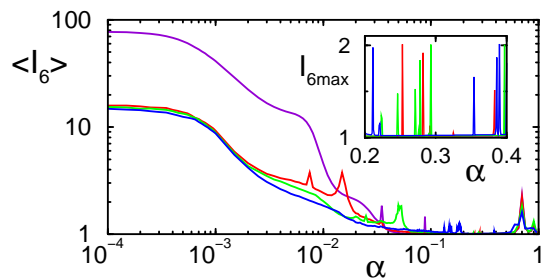}
\caption{(color) Many-particle localization for a chain of length $L=12$ with
$6$ excitations. The data refer to the first 12 sites of the chain
(\protect\ref{sequence}), the reduced bandwidth is $h/J=20$. The
purple, red, green, and blue curves give IPR for the coupling
parameter $\Delta=0,0.3,1$, and 3, respectively.
The inset shows the maximal $I_6$. Sharp isolated peaks for
$\Delta\neq 0$ result from the hybridization of many-particle on-site
states that are in resonance for the corresponding $\alpha$. The peaks
for $\Delta=0$ are due to the boundaries.}
\label{fig:many_loc}
\end{figure}

Numerical results on the many-particle IPR are shown in
Fig.~\ref{fig:many_loc}. We have studied chains of length $L=10, 12$,
and 14 with $L/2$ excitations, which have the largest number of states
for given $L$ ($\propto 2^L$ for large $L$). The results
were similar, and we present the data for $L=12$, in which case the
total number of states is 924.

For small $\alpha$, the IPR is independent of $\alpha$ and is large
because of the large number of resonating on-site states
$|\Phi_{n_1\ldots n_6}\rangle$. It is reduced by the interaction
$\propto J\Delta$ that splits the energy spectrum into subbands
depending on the number of occupied neighboring sites. On the whole,
the IPR decreases with increasing $\alpha$ as long as $\alpha \alt
0.4$. In the region $0.2\alt \alpha \alt 0.4$ we have $\langle
I_6\rangle \approx 1.01$ except for narrow peaks. This indicates that
away from the peaks all stationary states are strongly localized, as
confirmed by the data on $I_{6\max}=\max_{\lambda} I_{6\lambda}$.

A distinctive feature of the many-particle IPR as function of $\alpha$
are multiple resonant peaks, a part of which is resolved in
Fig.~\ref{fig:many_loc}. They occur when two on-site states resonate,
which gives $I_{6\max}\lesssim 2$. The strongest peaks come from
two-particle resonances. They happen when the two-particle energy
difference
\begin{equation}
\label{energy_difference}
\delta\varepsilon
=|\varepsilon_{k_1}+\varepsilon_{k_2}-
\varepsilon_{k_3}-\varepsilon_{k_4}|
\end{equation}
is close to $MJ\Delta$ with $M=0,1,2$.

As we increase $\alpha$ starting from $\alpha=0$, pronounced peaks of
$\langle I_6\rangle$ appear first for $\delta\varepsilon \approx
s\alpha h \approx J\Delta$ with $s=1,2$. They are due to hybridization
of pairs on sites $(n,n+1)$ and $(n,n+3)$ for $s=1$, and $(n,n+1)$ and
$(n-1,n+2)$ for $s=2$, for example ($\varkappa = 2$-transitions).

For larger $\alpha$, resonances occur when $s\alpha^nh\approx
MJ\Delta$ with $n\geq 2$. Such resonances require more intermediate
steps, with $\varkappa \geq 4$. The widths of the IPR peaks are small and
are in good agreement with simple estimates based on Eq.~(\ref{H_int})
\cite{elsewhere}. In between the peaks $I_{6\max} = 1.02$ for
$0.2\lesssim \alpha\lesssim 0.4$ and $h/J=20$.

A special role is played by two-particle resonances where $\delta
\varepsilon \ll J$ for all $\alpha < 0.4$. They emerge already for
$\varkappa = 2$-transitions $(n,n+1) \leftrightarrow (n-1,n+2)$. Here,
if $n$ and $n+2$ are prime numbers, $\delta \varepsilon\sim
\alpha^{n-1}h$ is extremely small for large $n$. Strong resonance
occurs for all $n=6k-1$, in which case $\delta \varepsilon/h\propto
\alpha^{\xi}$ with $\xi \geq 4$. Such $\delta \varepsilon$ is
unusually small for $\varkappa =2$. More many-particle resonances
happen for higher $\varkappa$. For different sections of the chain
(\ref{sequence}) we found that they could increase $\langle
I_6\rangle$ up to 1.15 between the peaks, for $h/J=20,0.2<\alpha<0.4$,
and $\Delta=1$. These resonances can be eliminated by modifying the
sequence (\ref{sequence}): $\varepsilon_n \to \varepsilon_n+\alpha^{\prime} h/2$
for $n=6k$. For appropriate $\alpha^{\prime} \sim 0.1$, this
modification brings $\langle I_6\rangle$ and $I_{6\max}$ back to
$\approx 1.01$ and $\approx 1.02$, respectively \cite{elsewhere,footnote2}.

We now outline an {\it alternative approach} to the problem of strong
localization, which is particularly relevant for quantum computing.
Qubit states have finite coherence time, estimated as $\lesssim
10^5 - 10^6J^{-1}$ for most models. It is sufficient to show that the
states remain localized for a time that exceeds the coherence
time. This will enable both gate operations, that take time $\sim
J^{-1}$, and measurement, that often requires more time.

Delocalization occurs through a transition to a resonating on-site
state. All resonant two-particle transitions up to a given number of
steps $\varkappa_0$ will be eliminated if, for $\varkappa <
\varkappa_0$, the on-site energy difference $\delta\varepsilon$
exceeds the maximum change of the interaction energy $\propto
J\Delta$. For $\varkappa_0=4$ this requires $J\Delta/h <
\alpha^2,\alpha^{\prime}/2, |\alpha-\alpha^{\prime}/2|$.  From
Eq.~(\ref{H_int}), the time needed for a transition with $\varkappa_0$
steps is $\agt K^{-\varkappa_{_0}}/J\Delta$, it scales as
$h^{\varkappa_{_0}}$.  Then if $\varkappa_0=4$, the lifetime of {\it
all} on-site states exceeds $J^{-1}$ by a factor $10^5$ for $h=30$,
$\alpha=0.3$, $\alpha^{\prime}\approx 0.2$, and $\Delta \alt 1$.

\begin{figure}[h]
\includegraphics[width=2.8in]{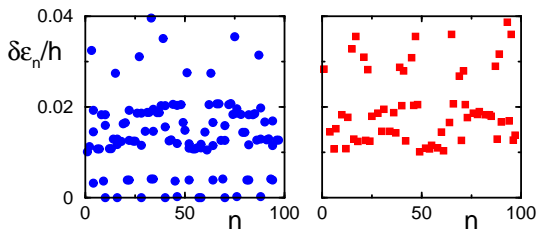}
\caption{(color online) The low-energy part of the two-particle energy
differences $\delta \varepsilon _n /h$
(\protect\ref{energy_difference}) for all transitions with
$\varkappa\leq 5$ in which one of the involved particles is 
on the $n$th site ($n> 2$). The data refer to $\alpha =0.25$. The left panel
corresponds to the sequence (\ref{sequence}).  The right
panel refers to the modified sequence with $\alpha^{\prime} = 0.22$
and shows the zero-energy gap.}
\label{fig:correction}
\end{figure}

The transition time is further dramatically increased if $\varkappa=4$
resonances are eliminated. The low-energy gap in $\delta\varepsilon$
for the corresponding transitions is $\sim\alpha^3h$ when $\alpha^{\prime} \gg
\alpha^3$. For example, for $\alpha = 0.25$ and $\alpha^{\prime} = 0.22$ this
gap is $\delta \varepsilon/h > 0.01$ for {\it all} sites $n>4$, cf.
Fig.~\ref{fig:correction}. This means that all states will remain
localized on their sites for a time $\agt K^6(J\Delta)^{-1}$, if
$2J\Delta/h < 0.01$.

In terms of the operation of a quantum computer, the on-site energy
sequence (\ref{sequence}) is advantageous, since one radiation
frequency can be used to resonantly excite different qubits.
Selective tuning to this frequency can be done without bringing
neighboring qubits in resonance with each other. In our approach,
localization does not require refocusing \cite{liquid_NMR}, which is
not always easy to implement. We avoid delocalization due to indirect
resonant $n\to n+2$ transitions, which undermines the approach
\cite{Benjamin03}. Compared to Ref.~\cite{Zhou02}, in our approach the
interaction does not have to be ever turned off, and no multi-qubit
encoding is necessary. In addition, our results are not limited to
systems with nearest neighbor coupling.

In conclusion, we have proposed a sequence of on-site energies that
leads to strong localization of single- and many-particle stationary
states of interacting spins or qubits. For an infinite chain, we
eliminate resonances between on-site states with the effective interstate
hopping integral up to $(J/h)^5$. This leads to a long lifetime of
localized many-particle states already for a comparatively narrow
energy bandwidth. The results show viability of scalable quantum
computers where the interqubit interaction is not turned off.

\begin{acknowledgments}
This research was supported in part by the Institute for Quantum
Sciences at Michigan State University and by the NSF through grant
No. ITR-0085922.
\end{acknowledgments}


\begin{thebibliography}{99}
\bibitem{footnote2} The obtained $\langle I_6\rangle-1$ were much
smaller than for a chain with random $\varepsilon_n$ uniformly
distributed over a band of the same width as (\ref{sequence}). On the
other hand, $\langle I_6\rangle$ for the sequence (\ref{sequence}) did
not increase when random terms up to $\sim h\alpha^4$ were added to
$\varepsilon_n$, which indicates that the  sequence (\ref{sequence}) is
robust with respect to errors.
\end{thebibliography}
\end{document}